\documentclass[english,preprint, aip,jcp,superscriptaddress,showpacs,floatfix]{revtex4-1}
\usepackage{amsmath}
\usepackage{graphicx}
\usepackage{makeidx}

\newcommand{\bra}[1]{\left \langle #1 \right \vert}
\newcommand{\ket}[1]{\left \vert #1 \right \rangle}
\newcommand{\braket}[2]{\langle #1 \vert  #2 \rangle}

\begin{document}

\title{Analysis of the Forward-Backward Trajectory Solution for the Mixed Quantum-Classical Liouville Equation}

\author{Chang-Yu Hsieh}
\affiliation{Chemical Physics Theory Group, Department of Chemistry, University of Toronto, Toronto, ON, M5S 3H6 Canada}

\author{Raymond Kapral}
\affiliation{Chemical Physics Theory Group, Department of Chemistry, University of Toronto, Toronto, ON, M5S 3H6 Canada}

\begin{abstract}
Mixed quantum-classical methods provide powerful algorithms for the simulation of quantum processes in large and complex systems. The forward-backward trajectory solution of the mixed quantum-classical Liouville equation in the mapping basis [J. Chem. Phys. 137, 22A507 (2012)] is one such scheme. It simulates the dynamics via the propagation of forward and backward trajectories of quantum coherent state variables, and the propagation of bath trajectories on a mean-field potential determined jointly by the forward and backward trajectories. An analysis of the properties of this solution, numerical tests of its validity and an investigation of its utility for the study of nonadiabtic quantum processes are given. In addition, we present an extension of this approximate solution that allows one to systematically improve the results. This extension, termed the jump forward-backward trajectory solution, is analyzed and tested in detail and its various implementations are discussed.
\end{abstract}

\maketitle


\section{Introduction} \label{sec:intro}

Nonadiabatic processes are important for the description of many chemical and biological processes such as proton and electron transfer reactions, vibrational relaxation, quantum reaction dynamics,  photochemical dynamics, and coherent energy transfer phenomena in biological systems~\cite{tully12}.  In all these processes, the dynamics of the system is strongly influenced by the environment in which it resides; hence, an accurate prediction of the system's properties entails the simulation of the system along with its environment.  To avoid the exponential growth in the computational costs of exact quantum simulations with the number of degrees of freedom, several mixed quantum-classical~\cite{0chap-tully98,kapral06_2} and semi-classical~\cite{herman94,stock05} schemes have been developed.

Simulation algorithms based on the quantum-classical Liouville equation~\cite{kapral06_2} have been used to study complex open quantum systems that display nonadiabatic phenomena~\cite{hanna05,hanna08}.  Differences among the algorithms often can be attributed to the basis set used to represent the quantum system, since the way the quantum degrees of freedom are propagated depends on basis set representations.  Entangled trajectories~\cite{donoso98,donoso03} can be used to simulate the quantum dynamics in the subsystem basis; surface-hopping-like algorithms~\cite{mackernan02,mackernan08} can be formulated in the adiabatic basis; the force basis~\cite{wan00,wan02}, obtained from diagonalizing the Hellmann-Feynman force, yields yet another propagation scheme. Another class of simple and efficient algorithms~\cite{kim-map08,hsieh12} can be constructed if the subsystem basis of an $N$-level quantum system is mapped onto the single-excitation space of $N$ fictitious harmonic oscillators. In this mapping representation~\cite{meyer79,stock97,thoss99} the discrete quantum degrees of freedom are mapped onto continuous variables, such as the positions and momenta of the fictitious oscillators. The Poisson bracket mapping equation~\cite{kim-map08,nassimi10,kelly12} and the forward-backward trajectory solution~\cite{hsieh12} are two such mapping-based approximate solution schemes for the quantum-classical Liouville equation. The forward-backward trajectory solution, the topic of this paper, is derived from the forward-backward form~\cite{nielsen01} of the formal solution of the mapping-transformed quantum-classical Liouville equation. Since the quantum time evolution operator is expressed in a coherent state basis, the time evolution of a mixed quantum-classical system is simulated through the forward and backward trajectory evolution of $N$ coherent state coordinates while the bath coordinates evolve under the influence of a mean potential that depends on these forward and backward trajectories. In this scheme, the quantum-classical Liouville dynamics is simulated via an ensemble of independent Newtonian trajectories.

In this paper we analyze various aspects of the forward-backward trajectory solution.  As shown earlier~\cite{hsieh12}, this solution satisfies the differential form of the quantum-classical Liouville equation and is formally invariant to the form (trace versus traceless) of the Hamiltonian. To derive such a continuous trajectory picture, an orthogonality approximation is made, which assumes that the coherent states connecting various time segments are orthogonal to each other. Here, we show that the formal invariance to the form of the Hamiltonian is broken due to the use of the orthogonality approximation. If the system-bath coupling is not a weak perturbation to the pure bath potential, the use of the trace-form Hamiltonian may yield poor results due to the possible presence of an inverted potential for the bath coordinates. To systematically improve the forward-backward trajectory solution, we generalize the simulation scheme by restricting the use of the orthogonality approximation to only certain time steps.  In the resulting new solution, the forward and backward coherent state trajectories experience discontinuous jumps in the phase space whenever the orthogonality approximation is not used; hence, the generalized solution is termed the jump forward-backward trajectory solution. Sampling of jumps is carried out by a Monte Carlo procedure. In order to improve the convergence, the focusing approximation~\cite{bonella03,dunkel08,huo12_jcp} is used in order to avoid performing the integrals over intermediate coherent state variables.

The outline of the paper is as follows: In Sec.~\ref{sec:theory} we summarize the principal elements of the forward-backward trajectory solution and present its generalization that involves jumps.  The focusing approximation is discussed in Sec.~\ref{sec:focusing}, and a simple picture of trajectory dynamics is given to illustrate the effects of focusing.  In Sec.~\ref{sec:models} we present simulation results on four test models, which are chosen to highlight important aspects of the algorithms.  The conclusions are given in Sec.~\ref{sec:con}.


\section{Forward-Backward Trajectory Solution} \label{sec:theory}
We consider a quantum subsystem coupled to a bath where the dynamics is described by the quantum-classical Liouville equation (QCLE).
The Hamiltonian has the form,
\begin{equation} \label{eq:ham}
\hat{H}_W(X)= H_b(X)+\hat{h}_s+\hat{V}_c(R) \equiv H_b(X)+\hat{h}(R),
\end{equation}
where the subscript $W$ refers to a partial Wigner transform over the bath degrees of freedom (DOF). Here $H_b(X)=P^2/2M+V_b(R)$ is the bath Hamiltonian
with $V_b(R)$ the bath potential energy, $\hat{h}_s=\hat{p}^2/2m+\hat{V}_s$ is the
subsystem Hamiltonian with $\hat{p}$ and $\hat{V}_s$ the subsystem momentum and
potential energy operators, and $\hat{V}_c(R)$ is the coupling potential
energy operator. The masses of the subsystem and bath particles are $m$ and $M$, respectively.

For a quantum operator $\hat{B}_W(X)$, which depends on the classical phase space variables $X=(R,P)=(R_1,R_2,...,R_{N_b},P_1,P_2,...,P_{N_b})$ of the bath, the formal solution of the QCLE is given by~\cite{kapral99},
\begin{equation}\label{bformalsol}
\hat{B}_W(X,t)=e^{i\hat{{\cal L}}t} \hat{B}_W(X),
\end{equation}
where the QCL operator is
\begin{equation}\label{eq:qclop-abs}
i\hat{{\mathcal L}} \hat{B}_W =\frac{i}{\hbar}[\hat{H}_W,\hat{B}_W] - \frac{1}{2}(\{\hat{H}_W,\hat{B}_W\}
-\{\hat{B}_W,\hat{H}_W\}),
\end{equation}
$[\hat{A}_W,\hat{B}_W]=\hat{A}_W\hat{B}_W-\hat{B}_W\hat{A}_W$ is the commutator, and $\{\hat{A}_W, \hat{B}_W \}=\frac{\partial \hat{A}_W}{\partial R}\frac{\partial \hat{B}_W }{\partial P} -  \frac{\partial \hat{A}_W}{\partial P}\frac{\partial \hat{B}_W}{\partial R}$ is the Poisson bracket with respect to $X$. We may also write the QCL operator as~\cite{nielsen01}
\begin{equation}
i\hat{{\mathcal L}} \hat{B}_W =\frac{i}{\hbar}\Big(
\stackrel{\rightarrow}{{\mathcal H}}_{\Lambda}\hat{B}_W-\hat{B}_W
\stackrel{\leftarrow}{{\mathcal H}}_{\Lambda}\Big),
\end{equation}
which involves the forward and backward evolution operators,
\begin{eqnarray}\label{eq:sah}
{\stackrel{\rightarrow}{{\cal H}}_{\Lambda}}=\hat{H}_W\left(1+ \frac{\hbar \Lambda}{2i}\right), \quad
{\stackrel{\leftarrow}{{\cal H}}_{\Lambda}}= \left(1+ \frac{\hbar \Lambda}{2i}\right)\hat{H}_W,
\end{eqnarray}
with $\Lambda$ the negative of the Poisson bracket operator, $\Lambda = \stackrel{\leftarrow}{\nabla}_P \cdot
\stackrel{\rightarrow}{\nabla}_R-\stackrel{\leftarrow}{\nabla}_R \cdot
\stackrel{\rightarrow}{\nabla}_P$. In this case the formal solution of the QCLE can be expressed as
\begin{equation}\label{bformalsol}
\hat{B}_W(X,t)= {\mathcal S}
\left(
e^{i{\stackrel{\rightarrow}{{\cal H}}_{\Lambda}}t/\hbar}\hat{B}_W(X)
e^{-i{\stackrel{\leftarrow}{{\cal H}}_{\Lambda}}t/\hbar} \right).
\end{equation}
The ${\mathcal S}$ operator in this form simply specifies the order in which products of the left and right operators act in order to be identical with the first form in Eq.~(\ref{eq:qclop-abs}). In particular, a general term ${\mathcal S}\left((\stackrel{\rightarrow}{{\mathcal H}}_{\Lambda})^j
\hat{B}_W (\stackrel{\leftarrow}{{\mathcal H}}_{\Lambda})^k\right)$ in the expansion of the exponential operators is composed of ${(j+k)!}/{j!k!}$ separate terms each with a pre-factor of ${j!k!}/{(j+k)!}$. Each of these separate terms corresponds to a specific order in which the
$\stackrel{\rightarrow}{{\cal H}}_{\Lambda}$ and $\stackrel{\leftarrow}{{\cal H}}_{\Lambda}$ operators act on $\hat{B}_W$.

\subsection{Hamiltonian dynamics in the coherent state phase space of the mapping representation}

We suppose that the time evolution of the quantum subsystem (coupled to the bath) can be accurately described within a truncated Hilbert space of dimension $N$.  Furthermore, the subsystem basis $\{ \ket{\lambda} \}$ is chosen for the matrix representations of quantum operators.

In the mapping representation, the state $|\lambda\rangle$ is replaced by $|m_{\lambda}\rangle$, an eigenfunction  of the Hamiltonian for $N$ fictitious harmonic oscillators~\cite{chap-schwinger65,stock05}, having occupation numbers which are limited to 0 or 1: $|\lambda\rangle\rightarrow|m_{\lambda}\rangle=|0_{1}, \cdots,1_{\lambda},\cdots0_{N}\rangle$. Creation and annihilation operators on these states, $\hat{a}_{\lambda}^{\dag}$ and $\hat{a}_{\lambda}$, satisfy the commutation relation $[ \hat{a}_\lambda, \hat{a}^{\dag}_{\lambda'}]   =  \delta_{\lambda,\lambda'}$. The actions of these operators on the single-excitation mapping states are $\hat{a}^{\dag}_\lambda \ket{0}  =  \ket{m_\lambda}$ and $\hat{a}_\lambda \ket{m_\lambda}  =  \ket{0}$, where
$\ket{0} = \ket{0_1 \dots 0_{N}}$ is the ground state of the mapping basis.

We may then define the mapping version of operators, $\hat{B}_{m}(X)$, given by $\hat{B}_{m}(X)=B_{W}^{\lambda\lambda'}(X)
\hat{a}_{\lambda}^{\dag}\hat{a}_{\lambda'},$
such that matrix elements of $\hat{B}_W$ in the subsystem basis are equal to the matrix elements of the corresponding mapping operator:
$B_{W}^{\lambda\lambda'}(X) =\langle\lambda|\hat{B}_{W}(X)|\lambda'\rangle=\langle
m_{\lambda}|\hat{B}_{m}(X)|m_{\lambda'}\rangle$. (The Einstein summation convention will be used throughout although sometimes sums will be explicitly written if there is the possibility of confusion.)  In particular, the mapping Hamiltonian operator is
\begin{equation}
\label{eq:mapping_hamiltonian}
\hat{H}_m= H_b(X)+  h^{\lambda \lambda'}(R) \hat{a}^{\dag}_\lambda \hat{a}_{\lambda'}\equiv H_b(X)+\hat{h}_m,
\end{equation}
where we applied the mapping transformation only on the part of the Hamiltonian that involves
the subsystem DOF in Eq.~(\ref{eq:mapping_hamiltonian}). The pure bath term, $\hat{H}_{b}(X)$ in
Eq.~(\ref{eq:ham}), acts as an identity operator in the subsystem basis and is mapped onto the
identity operator of the mapping space. The formal solution in Eq.~(\ref{bformalsol}) may be expressed in mapping operators and is given by
\begin{equation}\label{eq:bformalsol-map}
\hat{B}_m(X,t)={\mathcal S} \left(
e^{i{\stackrel{\rightarrow}{{\cal H}^m_{\Lambda}}}t/\hbar}\hat{B}_m(X)
e^{-i{\stackrel{\leftarrow}{{\cal H}^m_{\Lambda}}}t/\hbar} \right),
\end{equation}
where $\stackrel{\rightarrow}{{\mathcal H}^m_{\Lambda}}$ is given by ${\stackrel{\rightarrow}{{\cal H}^m_{\Lambda}}}=\hat{H}_m(1+ \hbar \Lambda/2i)$,
with an analogous definition for ${\stackrel{\leftarrow}{{\cal H}^m_{\Lambda}}}$.

We now define the coherent states $\ket{z}$ in the mapping space,
$\hat{a}_{\lambda} \ket{z}= z_\lambda \ket{z}$ and  $\bra{z}\hat{a}_{\lambda}^\dagger= z_\lambda^* \bra{z}$,
where $\ket{z}=\ket{z_1,\dots,z_N}$ and the eigenvalue is $z_\lambda= (q_\lambda + i  p_\lambda )/\sqrt{2\hbar}$. The variables $q=(q_1, \dots, q_{N})$ and $p=(p_1, \dots, p_{N})$ are mean coordinates and momenta of the harmonic oscillators in the state $\ket{z}$, respectively; i.e.,
we have $\bra{z} \hat{q}_\lambda \ket{z} = q_\lambda$ and $\bra{z} \hat{p}_\lambda \ket{z} = p_\lambda$.
The coherent states form an overcomplete basis with the inner produce between any two such states, $\bra{z}z' \rangle = e^{-\frac{1}{2}(|z-z'|^2) -i/2 (z\cdot z^{\prime *}-z^{*}\cdot z^{\prime})}$.  Finally, we remark that the coherent states provide the resolution of identity,
\begin{equation}\label{eq:coh_resol}
1= \int \frac{d^2z}{\pi^{N}} \ket{z}\bra{z},
\end{equation}
where $d^2z=d(\Re(z)) d(\Im(z))=dq dp/(2\hbar)^N$.

Following our earlier work~\cite{hsieh12}, we then decompose the forward and backward evolution operators in Eq.~(\ref{eq:bformalsol-map}) into a concatenation of $M$ short-time evolutions with $\Delta t_i = \tau$ and $M\tau = t$.  In each short-time interval $\Delta t_i$, we introduce two sets of coherent states, $\ket{z_i}$ and $\ket{z^{\prime}_i}$, via Eq.~(\ref{eq:coh_resol}).  After some algebra, the matrix elements of Eq.~(\ref{eq:bformalsol-map}) can be approximated by
\begin{eqnarray}\label{eq:mqc_soln-mat2}
&&{B}^{\lambda \lambda'}_W (X,t) =\sum_{\mu \mu'}\int \prod_{i=1}^M \frac{d^2z_i}{\pi^{N}} \frac{d^2z'_i}{\pi^{N}}
\langle m_{\lambda} \ket{z_1} \bra{z^\prime_1} m_{\lambda'}\rangle \nonumber \\
&& \qquad \quad
e^{i{\mathcal L}_e(X,z_1,z_1') \Delta t_1} \Big(
\braket{z_1(\Delta t_1)}{z_2}\nonumber \\
&& \qquad \quad
    e^{i{\mathcal L}_e(X,z_2,z_2') \Delta t_2} \Big( \braket{z_2(\Delta t_2}{z_3} \dots \ket{z_M}\nonumber \\
&& \qquad \quad
e^{i{\mathcal L}_e(X,z_M,z_M') \Delta t_M} \Big( \braket{z_M(\Delta t_M}{m_{\mu}} \nonumber \\
&& \qquad \quad {B}^{\mu \mu'}_W(X) \braket{m_{\mu'}}{z^{\prime}_M(\Delta t_M)}\Big)  \\
&& \qquad \quad \bra{z^\prime_M}  \dots
\ket{z^\prime_2(\Delta t_2)} \Big)
\braket{z^\prime_2}{z^\prime_1(\Delta t_1)}  \Big).\nonumber
\end{eqnarray}
Equation (\ref{eq:mqc_soln-mat2}) is evaluated sequentially, from smallest to largest times, by taking the bath phase space propagators in expressions such as $e^{i{\mathcal L}_e(X,z_i,z_i') \Delta t_i} ( \cdots )$ to act on all quantities in the parentheses, including other propagators at later times, $\Delta t_{j>i}$. The bath phase space propagator reads
$i {\mathcal L}_{e}(X,z,z')=\frac{P}{M}  \cdot \frac{\partial}{\partial R} -\frac{\partial V_e(X,z,z')}{\partial R}  \cdot \frac{\partial}{\partial P}$ with
\begin{eqnarray}\label{eq:hamil_pot}
V_e(X,z,z') & = & V_b(R) -\sum_{\lambda}h^{\lambda\lambda}+\frac{h^{\lambda\lambda^{\prime}}}{2}\left(z^*_\lambda z_{\lambda^{\prime}} + z^{\prime *}_{\lambda}z^{\prime}_{\lambda^{\prime}}\right), \nonumber \\
& \equiv & V_0(R) +\frac{1}{2}\left(h^{\lambda\lambda^{\prime}}z^*_\lambda z_{\lambda^{\prime}} + h^{\lambda\lambda^{\prime}}z^{\prime *}_{\lambda}z^{\prime}_{\lambda^{\prime}}\right), \\
& \equiv & V_0(R) + \frac{1}{2} \left( h_{cl}(R,z) + h_{cl}(R,z') \right). \nonumber
\end{eqnarray}
In the first two lines, we suppress the $R$-dependence of the matrix elements $h^{\lambda\lambda^{\prime}}(R)$. Similar to other QCLE solutions, the bath dynamics are governed by classical trajectories in the bath phase space.

To obtain Eq.~(\ref{eq:mqc_soln-mat2}), we also use the fact the coherent state evolution under a quadratic Hamiltonian $\hat{h}_m(R)$ can be exactly evaluated by $ e^{-i\hat{h}_m \frac{\Delta t_i}{\hbar}}\ket{z}= \ket{z(\Delta t_i)}$, where the trajectory evolution of $z_\lambda$ is governed by $\frac{dz_{\lambda}}{dt} =-\frac{i}{\hbar}\frac{\partial h_{cl}(R,z)}{\partial{z}^*_{\lambda}}$. Thus the quantum subsystem dynamics is governed by segments of coherent state trajectories (defined by $z_i$ and $z^{\prime}_i$), which are not necessarily continuous across the time steps because the coherent state variables $z_i$ and $z_{i+1}$ are independent to each other. By making the orthogonality approximation to the inner products, $\langle z_i(t_i) \ket{z_{i+1}} \approx \pi^{N}\delta(z_{i+1}-z_i(t_i))$, we construct a smooth trajectory solution of the quantum dynamics.  Furthermore, by scaling the coherent state variables $z_{\text{scaled}}  = z/\sqrt{2}$, we also show that the trajectories in the scaled phase space $(X,z_{\text{scaled}},z^{\prime}_{\text{scaled}})$ are strictly Newtonian trajectories.  In the rest of the article, we will consistently use scaled coherent state variables and drop the subscript ``scaled".

In summary, the orthogonality approximation and the scaling of coherent state variables helps to simplify Eq.~(\ref{eq:mqc_soln-mat2}) and its interpretation. We have
\begin{eqnarray}\label{eq:mqc_soln-mat6}
&&{B}^{\lambda \lambda'}_W (X,t) =\sum_{\mu \mu'}\int dx dx^{\prime} \phi(x) \phi(x') \nonumber \\
&& \qquad \times \frac{1}{\hbar}  (q_{\lambda}+ip_{\lambda})(q'_{\lambda^\prime}-ip'_{\lambda^\prime}) {B}^{\mu \mu'}_W(X_{t})\nonumber \\
&& \qquad \times \frac{1}{\hbar} (q_{\mu}(t)-ip_{\mu}(t))(q'_{\mu^\prime}(t)+ip'_{\mu^\prime}(t)),
\end{eqnarray}
where $x = (q, p)$, $dx = dqdp$, and $\phi(x)=\left(\hbar\right)^{-N}e^{-\sum_\nu (q_\nu^2+p_\nu^2)/\hbar}$ is the normalized Gaussian distribution function.
We have written this equation in the scaled coherent state variables $z_\lambda = (q_\lambda + i p_\lambda)/\hbar$.
The trajectories of $X_t$, $x_t$ and $x^{\prime}_t$ are governed by Hamilton's equations,
\begin{eqnarray}\label{eq:vol-nonH}
&&\frac{d q_\mu}{dt}= \frac{\partial  H_{e}(X,x,x')}{\partial p_\mu}, \quad \frac{d p_\mu}{dt}= -\frac{\partial H_{e}(X,x,x')}{\partial q_\mu} \nonumber \\
&&\frac{d q'_\mu}{dt}= \frac{\partial H_{e}(X,x,x')}{\partial p'_\mu}, \quad \frac{d p'_\mu}{dt}= -\frac{\partial H_{e}(X,x,x')}{\partial q'_\mu}\nonumber \\
&&\frac{dR}{dt}=\frac{P}{M}, \qquad  \frac{dP}{dt}=- \frac{\partial H_{e}(X,x,x')}{\partial R},
\end{eqnarray}
where $H_{e}(X,x,x')=P^2/2M + V_0(R) + \frac{h^{\lambda\lambda'}(R)}{2\hbar}(q_\lambda q_{\lambda'}+ p_\lambda p_{\lambda'}+q'_\lambda q'_{\lambda'}+ p'_\lambda p'_{\lambda'})$.  In the following discussion, we take $\hbar = 1$ for simplicity.

Similar to the trajectory solution of the Poisson bracket mapping equation (PBME), the FBTS approximates the quantum-classical dynamics in terms of an ensemble of independent Newtonian trajectories. However, the fact that the quantum-related phase space in the FBTS is twice as large as the PBME phase space allows more complex evolutions of these trajectories and more accurate characterization of the bath potential surface.  In Sec.~\ref{sec:fic}, we will discuss the reduction of FB trajectories to PBME-like trajectories under some specific conditions.

\subsection{Jump  forward-backward trajectory solution}\label{sec:jfb}

As discussed in Ref.~[\onlinecite{hsieh12}], Eq.~(\ref{eq:mqc_soln-mat2}) yields exact QCLE dynamics if one does not impose the orthogonality approximation on the coherent states.  However, the large number of intermediate variables $z_i$ ($z^{\prime}_i$) precludes practical computations of the integrals involving these variables.  To extract QCLE results from FB trajectories, a systematic method is needed which can make an efficient compromise between computational costs and convergence.  One possible method to achieve this compromise in a controlled manner is to select a subsequence of $K$ time steps $\{t_{i_{v}}, v=1,\dots,K\}$ in Eq.~(\ref{eq:mqc_soln-mat2}) and evaluate the integrals of $z_{i_{v}}$ and $z^{\prime}_{i_{v}}$ explicitly (\textit{i.e.} without resorting to the orthogonality approximations).  According to this prescription, the continuous FB trajectories experience $K$ discontinuous jumps in the $(x,x')$ phase space.

In practice, this scheme can be carried out in several ways depending on the selection of the $K$ time steps.  In the simplest case, one may select every $(M/K)$ time steps from a total of $M$ steps to fully evaluate the coherent state integrals in Eq.~(\ref{eq:mqc_soln-mat2}):
\begin{eqnarray}\label{eq:mqc_soln-jump}
&&{B}^{\lambda \lambda'}_W (X,t) =\sum_{\mu \mu^{\prime}}
\sum_{\substack{s_0 s'_0 \dots \\ s_{K-1} s'_{K-1}}}
\int \prod_{v=0}^{K} dx_v dx^{\prime}_v\phi(x_v) \phi(x'_v)  \nonumber \\
&& \times  (q_{0\lambda}+ip_{0\lambda})(q'_{0\lambda^\prime}-ip'_{0\lambda^\prime})B^{\mu \mu'}_W(X_{t})  \\
&& \times \left\{ \prod_{v=1}^{K}
\left(q_{(v-1)s_{v-1}}(\tau_{v})-ip_{(v-1)s_{v-1}}(\tau_{v})\right)
\left(q_{vs_{v}}+ip_{vs_{v}}\right) \right\} \nonumber \\
&& \times \left\{ \prod_{v=1}^{K}
\left(q'_{(v-1)s_{v-1}}(\tau_{v})+ip'_{(v-1)s_{v-1}}(\tau_{v})\right)
\left(q'_{vs_{v}}-ip'_{vs_{v}}\right) \right\} \nonumber \\
&& \times  (q^{}_{K\mu}(\tau_{K+1})-ip^{}_{K\mu}(\tau_{K+1}))(q'_{K\mu^\prime}(\tau_{K+1})+ip'_{K\mu^\prime}(\tau_{K+1})), \nonumber
\end{eqnarray}
where the subscripts, $v$ and $s$, refer to the $v$-th time step and the $s$-th component of $q$ and $p$ vectors, respectively, and $\tau_v = t_{i_v} - t_{i_{v-1}}$ with $t_{i_0} = 0$ and $t_{i_{K+1}} = t$.  To obtain Eq.~(\ref{eq:mqc_soln-jump}), we  inserted a projection operator $\mathcal{P} = \sum_s \ket{m_s}\bra{m_s}$ between the inner product of two coherent states at every selected time step,
$\langle z_i(\tau_{i+1}) \vert \mathcal{P} \vert z_{i+1} \rangle$.
In Appendix A, we show that the insertions of $\mathcal{P}$ do not introduce further approximations in this calculation.  It is desirable to insert $\mathcal{P}$ every time we introduce a discontinuous jump to the trajectories, since then $\phi(x_i)$ and $\phi(x'_i)$ appear at each time step.  Otherwise, displaced Gaussian functions must be considered. The current form of the equation is also more suitable for introducing the focusing, an approximation scheme which is discussed in the next section.

A stochastic generalization of the fixed-interval selection method can also be used.  We may write
\begin{eqnarray}\label{eq:mqc_soln-jump2}
&&{B}^{\lambda \lambda'}_{\kappa_1,\dots,\kappa_{L}} (X,t) =\sum_{\mu \mu^{\prime}}
\sum^{\prime}_{\substack{s_0 s'_0 \dots \\ s_{L-1} s'_{L-1}}}
\int \prod_{v=0}^{L} dx_v dx^{\prime}_v \nonumber \\
&& \times \psi_{\kappa_v}(x_v) \psi_{\kappa_v}(x'_v)   (q_{0\lambda}+ip_{0\lambda})(q'_{0\lambda^\prime}-ip'_{0\lambda^\prime})B^{\mu \mu'}_W(X_{t})  \\
&& \times \left\{ \prod_{v}^{\prime}
\left(q_{(v-1)s_{v-1}}(\tau_{v})-ip_{(v-1)s_{v-1}}(\tau_{v})\right)
\left(q_{vs_{v}}+ip_{vs_{v}}\right)  \right\} \nonumber \\
&& \times \left\{ \prod_{v}^{\prime}
\left(q'_{(v-1)s_{v-1}}(\tau_{v})+ip'_{(v-1)s_{v-1}}(\tau_{v})\right)
\left(q'_{vs_{v}}-ip'_{vs_{v}}\right) \right\} \nonumber \\
&& \times  (q^{}_{L\mu}(\tau_{L+1})-ip^{}_{L\mu}(\tau_{L+1}))(q'_{L\mu^\prime}(\tau_{L+1})+ip'_{L\mu^\prime}(\tau_{L+1})), \nonumber
\end{eqnarray}
where $\kappa_v = 0 \text{ or } 1$.  A prime (${}^{\prime}$) is put on the second summation symbol and the last two product symbols in the equation to imply that the summations and multiplications only exist for a pair of $s_v$ and $s'_v$ if $\kappa_v = 1$.  Furthermore, we replace the normalized Gaussian functions $\phi(x_v)$ with
\begin{equation}\label{eq:mqc_stochastic_fn}
\psi_{\kappa_v}(x_v) = \left\{\begin{array}{ll}
	 \delta(x_v-x_{v-1}(\tau_v)), & \kappa_v = 0, \\
	\phi(x_v), & \kappa_v = 1.
\end{array}\right.
\end{equation}
The interpretation of Eqs.~(\ref{eq:mqc_soln-jump2}) and~(\ref{eq:mqc_stochastic_fn}) is straightforward.  We select $L$ time steps that are $J$ steps apart, $\it{i.e.}$ $JL=M$.
For a given binary sequence of $\{\kappa_1 ,\dots, \kappa_L\}$, we fully evaluate the coherent state integrals at the $vJ$-th time step if $\kappa_v=1$.  For other selected time steps corresponding to $\kappa_v=0$, we still apply the orthogonality approximation to simplify the calculation.  Finally, we remark that matrix element of interest, $B^{\lambda\lambda'}_W(X,t)$, can be obtained as an average of all possible binary sequences of $\kappa$,
\begin{eqnarray}\label{eq:mqc_soln-jump-avg}
B^{\lambda\lambda'}_W(X,t) &=& \sum_{\kappa_1,\dots \kappa_L} P_{\{\kappa\}}{B}^{\lambda \lambda'}_{\kappa_1,\dots,\kappa_{L}} (X,t),
\end{eqnarray}
where $P_{\{\kappa\}}$ denotes the discrete probability distribution of a given binary sequence of $\{\kappa_1 ,\dots, \kappa_L\}$.  Equation (\ref{eq:mqc_soln-jump-avg}) reduces to Eq.~(\ref{eq:mqc_soln-jump}) if one takes $J=M/K$ and $P_{\{\kappa_v=1\}}=1$.  Also, Eq.~(\ref{eq:mqc_soln-jump-avg}) reduces to Eq.~(\ref{eq:mqc_soln-mat6}) if one takes $P_{\{\kappa_v=0\}}=1$ .

We define the jump forward-backward trajectory solution (JFBTS) by Eqs.~(\ref{eq:mqc_soln-jump2}) - (\ref{eq:mqc_soln-jump-avg}).  In particular, we use the binomial distribution for $P_{\{\kappa\}}$ in this study.  We take the Bernoulli trial probability $p =\frac{K}{M/J}$ where $K$ is the average number of jumps (equivalent to the average number of 1's in the binary sequence of $\kappa$).  Equation~(\ref{eq:mqc_soln-jump-avg}) also has a simple interpretation in terms of trajectories if the integrals are evaluated using Monte Carlo (MC) sampling. During the propagation of a trajectory, a uniformly-distributed random number $\xi$ is generated at every $J$-th step.  If $\xi < p$, then we introduce a jump; otherwise, we continue to propagate the continuous trajectory until the next $J$-th step.

Finally, we note the close resemblance in structure between the JFBTS and IPLDM ~\cite{huo12_jcp} (iterative partial linearized density matrix), a similar generalization scheme that is applied to the PLDM algorithm in order to improve the results.  The differences between JFBTS and IPLDM can mostly be attributed to the differences between FBTS and PLDM~\cite{huo11, huo12_jcp,hsieh12}.

\section{Further approximation: Focusing} \label{sec:focusing}

In this section, we provide a detailed discussion of the effects of focusing, an approximation proposed\cite{bonella03} by Bonella \textit{et al.}, on the FBTS and JFBTS. Focusing can be defined as follows,
\begin{eqnarray}\label{eq:focusing}
&& \int dx \, \phi(x)  (q_\lambda^2+p_\lambda^2)f(x) \nonumber \\
&& \quad \approx
\int dx \, \delta(q_\lambda^2+p_\lambda^2-1)
\prod_{\mu \neq \lambda} \delta(q_\mu)\delta(p_\mu) \nonumber \\
&& \qquad \times (q_\lambda^2+p_\lambda^2)f(x),
\end{eqnarray}
where $f(x)$ should be at most weakly dependent on $x$ variables.
In order to obtain the approximate integral, one defines the mapping weight $\rho(x)=(q_\lambda^2+p_\lambda^2)\phi(x)$, and assumes that only the maximum points, satisfying the Dirac delta functions in Eq.~(\ref{eq:focusing}), of this weight contribute significantly to the integral.

Equation (\ref{eq:focusing}) may not seem to be directly related to any integrals we have presented so far.  However, if one considers the integral in Eq.~(\ref{eq:mqc_soln-mat6}) with $t \approx 0$, then the integrand
$\sum_{\mu\mu'}B^{\mu\mu'}_W(X_t)(q_\mu(t)-ip_\mu(t))(q^{\prime}_{\mu^{\prime}}(t)+ip^{\prime}_{\mu^{\prime}}(t))$ can be approximated by a single term (where
$\mu=\lambda$ and $\mu'=\lambda'$) of the summation.  Furthermore, near the initial time $t=0$, the integrand $B^{\lambda\lambda'}_W(X_t)$ is weakly dependent on coherent state variables, and the polynomial factors involving $x$ and $x'$ can be approximated by their initial values.  Same arguments are also applicable to Eq.~(\ref{eq:mqc_soln-jump2}). Hence, focusing can  approximate the integrals of Eqs.~(\ref{eq:mqc_soln-mat6}) and (\ref{eq:mqc_soln-jump2}) fairly well when $t$ is small or the system-bath coupling is weak; {\it i.e.}, a separation of time scale can be assumed. Unfortunately, these assumptions are not valid in the long-time limit for most physical systems.

By using focusing, one avoids the full sampling from the Gaussians in the MC evaluation of Eqs.~(\ref{eq:mqc_soln-mat6}) and (\ref{eq:mqc_soln-jump2}). As indicated in many numerical tests\cite{dunkel08,bonella05,kim-map08,huo12_jcp}, focusing often yields a converged result with at least an order of magnitude fewer trajectories than that required for a full calculations.  However, when used indiscriminately, focusing might yield poor results~\cite{huo12_jcp}.

\subsection{Focused Initial Condition in the FBTS} \label{sec:fic}
We suppose that the quantum subsystem is initially in a single subsystem state $\ket{\lambda}$. This requirement is not as restrictive as it might appear, since our formalism does not dictate which particular quantum subsystem basis is to be used.  Therefore, we can often take the initial state of the system to be a vector in the basis. When focusing~\cite{bonella05} is applied to the FBTS, the approximation restricts the initial conditions of the trajectories to a very narrow region in the phase space specified by the Dirac delta functions in Eq.~(\ref{eq:focusing}).

First, we discuss the effects of the focused initial conditions on the forward and backward coherent state trajectories.  In Appendix B, we prove that the backward coherent state trajectories can be replaced by the forward trajectories in the FBTS if focused initial conditions are imposed and the required assumptions (such as the separation of time scales) are valid.  This observation shows that the two sets of trajectories are highly synchronized in the sense explained in Appendix B. One manifestation of such synchronization is the reduction of noise in the calculation of the phase factor  of the quantum subsystem density matrix element, $ \rho^{\mu\mu'}_s(X, t)$.  In the full FBST calculation, such a matrix element is determined by the ensemble average of $(q_\lambda+ip_\lambda)(q'_\lambda-ip'_\lambda)(q_\mu(t)-ip_\mu(t))(q'_{\mu'}(t)+ip'_{\mu'}(t))$ according to Eq.~(\ref{eq:mqc_soln-mat6}).  However, in the case of focused initial conditions, one can simply replace $(q',p')$ with $(q,p)$ in the above expression for the ensemble average.  For instance, if one employs the formalism to compute a diagonal density matrix element, $\rho^{\mu\mu}(X, t)$, then one finds that this replacement yields an exact real number as required.  For the full calculation, one needs to average over a large number of trajectories to suppress the noisy imaginary components.

Next we consider the effects of focused initial conditions on bath trajectories. As mentioned earlier, focusing should be a valid approximation for an initial  short time interval or for a weak subsystem-bath coupling. We will take this into account by assuming that spatial derivatives of
${\bf C}(R)$, which is a column matrix of eigenvectors of $\hat{h}(R)$, can be neglected. Under this assumption, the bath momenta evolve according to,
\begin{eqnarray}\label{eq:focus_bathp}
 \frac{dP}{dt} & = &
-\frac{\partial V_0(R)}{\partial R}
-\frac{\partial h^{\nu\nu'}(R)}{\partial R}(q_\nu(t) q_{\nu'}(t)
+ p_\nu(t) p_{\nu'}(t)) \nonumber \\
& = &  -\frac{\partial V_0(R)}{\partial R}
-\sum_{\chi\mu\mu'\nu\nu'}C^{-1}_{\chi \mu}C_{\mu \nu}\frac{\partial h^{\nu\nu'}(R)}{\partial R} C_{\nu' \mu'}C^{-1}_{\mu' \chi}  \nonumber \\
&& \quad \times \cos(\Delta \omega_{\mu\mu'} t) \left( q_\chi^2 + p_\chi^2 \right) \nonumber \\
& = &  -\frac{\partial V_0(R)}{\partial R}
-\sum_{\mu\mu'\nu\nu'}C^{-1}_{\lambda \mu}C_{\mu \nu}\frac{\partial h^{\nu\nu'}(R)}{\partial R} C_{\nu' \mu'}C^{-1}_{\mu' \lambda} \nonumber \\
&& \times \cos(\Delta \omega_{\mu\mu'}t),
 \end{eqnarray}
where $\Delta \omega_{\mu\mu'} = \omega_\mu - \omega_{\mu'}$. In writing this equation, as discussed above, we replaced the backward coherent state variables with the forward ones. To obtain the second equality, the equation of motion for the coherent state variables was integrated to yield analytical expressions for $q(t)$ and $p(t)$. The final form of this equation was obtained using the focused initial condition $q_\chi^2+p_\chi^2 = \delta_{\chi\lambda}$.  In the full calculation, the system-bath coupling term (the second equality in the equation) should be averaged over all possible values of $\sum_{\chi}(q_{\chi}^2+p_{\chi}^2) = r$ with respect to the Gaussian distribution $\phi(r)$.

For situations typical of a scattering problem, where the quantum subsystem is initialized in the asymptotic region with $C_{\mu\nu} = \delta_{\mu\nu}$ (\text{i.e.} the subsystem basis and the adiabatic basis coincide), the final form of Eq.~(\ref{eq:focus_bathp}) reduces to $dP/dt = -\partial V_0(R)/\partial R -\partial h^{\lambda\lambda}(R)/\partial R$. This simple bath dynamics is the adiabatic approximation in the asymptotic region.  In contrast, the full calculation requires summation over a large number of trajectories such that the interferences among them leads to the adiabatic approximation. However, in the interaction region of the scattering potential, this adiabatic approximation is certainly not sufficient.  Even though Eq.~(\ref{eq:focus_bathp}) shows that focusing includes more than the adiabatic approximation on the bath dynamics, it simply does not include enough corrections to fully describe the dynamics in the interaction region  in many cases.  For instance, in a recent calculations on a simple avoided crossing model~\cite{Tully90}, Huo and Coker showed~\cite{huo11}  that the correct bath properties cannot be reliably extracted from focused initial conditions.

\subsection{Iterative Focusing in the JFBTS}\label{sec:if}
While focusing is not an essential part of the FBTS, it is often the only tractable means by which one can obtain a converged JFBTS solution when one needs more than five jumps in order to obtain the QCLE result.  Similar to several iterative schemes of partially linearized path integral formalisms\cite{dunkel08,huo12_jcp}, such as IPLDM, we rely on focusing's superior convergence properties to minimize the increasing number of trajectories required to produce a stable result.

Comparing Eqs.~(\ref{eq:mqc_soln-mat6}) and (\ref{eq:mqc_soln-jump2}), the JFBTS essentially adds a sequence of summations over intermediate quantum states, $s_{i}$ and $s'_{i}$, and integrals over coherent state variables $z_i$ and $z'_i$ on top of Eq.~(\ref{eq:mqc_soln-mat6}). While an exact evaluation of Eq.~(\ref{eq:mqc_soln-jump2}) is conceptually straightforward, the numerical implementation incorporating the focusing approximation is slightly more complicated.  First we evaluate the sequence of intermediate summations by an importance sampling MC calculation.  This technique has been adopted in several related QCLE and linearized path integral algorithms\cite{mackernan08,dunkel08,huo12_jcp}.  Thus, in each MC run, we randomly select a pair of forward and backward quantum states $(s_i, s'_i)$ at every $i$-th jump instead of summing over all quantum states.  The pair selection is governed by a discrete probability distribution,
$P_{s_{i},s^{\prime}_{i}} = (q_{s_i}^2+p_{s_i}^2)(q^{\prime 2}_{s^{\prime}_{i}}+p^{\prime 2}_{s^{\prime}_{i}})/\mathcal{N}_i$ with
$\mathcal{N}_i = \sum_{s_i,s'_i} (q_{s_i}^2+p_{s_i}^2)(q^{\prime 2}_{s^{\prime}_{i}}+p^{\prime 2}_{s^{\prime}_{i}})$.  This probability distribution is based on the intuition that the  selection of the pair indices should be related to the present distribution of the population weight, $(q^2+p^2)$, over all quantum states.  A re-weighting factor, $1/P_{s_{i},s^{\prime}_{i}}$, is added to the MC evaluation whenever such an importance sampling is made.
Once the selection of a pair of quantum indices is made for the $i$-th jump, we use the focusing approximation with the condition $\delta(q_{s_i}^2+p_{s_i}^2-1) \delta(q_{s'_i}^{\prime 2}+p_{s'_i}^{\prime 2}-1)$ to evaluate the integral over
$z_i$ and $z'_i$.  In the above focusing condition, we imply that all unspecified components of forward and backward variables are set to zero.

Now we analyze the effects of iterative focusing on the trajectories in the JFBTS formalism.  First, we provide an intuitive picture of the trajectory evolution when full sampling over coherent state variables and full summation over quantum states at each intermediate step are performed exactly.  At the initial time, $t=0$, we assign initial conditions of $q_\nu$ and $p_\nu$ for each quantum state $\ket{m_\nu}$ by sampling from the Gaussian $\phi(x)$.  The trajectories are then propagated with Hamiltonian dynamics based on Eq.~(\ref{eq:vol-nonH}).  At each time step, we stochastically decide if a jump should be introduced.  When a jump is introduced at time $t$, we store a snapshot of the trajectories by saving the polynomial factor $(q_\nu(t) \pm ip_\nu(t))$ for each $\nu$, where the plus and minus signs are for forward and backward coherent state variables, respectively.  Then we discard the trajectories and re-sample new initial conditions for those trajectories.  We also take a snapshot of the new trajectories by saving the polynomial factor $(q_\nu \mp ip_\nu)$ prior to further propagation.  The same process is then be repeated until the simulated time length has been reached.  Due to the interferences in the full summations of the products of the polynomial factors in Eq.~(\ref{eq:mqc_soln-jump2}), only those trajectories with rather minimal discontinuities contribute significantly to the final result.

Next we describe a rather different evolution of trajectories in the phase space. The combination of the importance sampling of intermediate quantum state indices and the focusing approximation for the coherent state integrals yields a picture involving trajectory collapse whenever a jump is introduced.  In this numerical implementation, each forward and backward trajectory can only be initialized in the given pair of quantum states, $\ket{m_{s_i}}$ and $\ket{m_{s'_i}}$.  Once the initial conditions are assigned to the trajectories, the population weights, $q^2_s+p^2_s$, of the trajectories will gradually shift from the initial states,  $\ket{m_{s_i}}$ and $\ket{m_{s'_i}}$, to other states in the forward and backward phase space, respectively.  After some time, the population weights may become more uniformly distributed over all available states.  There will then come a moment where a jump is introduced, a new pair of quantum indices is selected, and a new focusing condition is imposed to collapse all trajectories onto the new pair of forward and backward states.

These two rather different pictures of the trajectory evolution make one wonder if iterative focusing applied to the JFBTS can be trusted. We argue that the result should be reliable.  Different from the focused initial condition where the states to be focused onto are fixed, iterative focusing collapses the forward and backward trajectories into all possible combinations of pairs of quantum states. Therefore, if we have a large enough MC sample, we will eventually obtain collapsed trajectories re-initialized in all possible quantum states and recover an ensemble picture that is in accord with that of the full-sampling case.


\section{Simulation results}\label{sec:models}
We now present the results obtained from simulations of the FBTS and JFBTS for four models, which are selected to examine different aspects of the simulation algorithms.  In particular, we use these results to illustrate the following: First, the differential form of the FBTS has been shown~\cite{hsieh12} to be invariant to the following forms of Hamiltonian,
\begin{eqnarray}\label{eq:2formH}
\hat{H}_W(x) = \left \{
\begin{array}{ll}
H_b(X) + \hat{h}(R), & \text{trace form,} \\
& \\
\overline{H}_b(X) + \overline{h}(R), & \text{traceless form,}
\end{array}
\right.
\end{eqnarray}
where $\overline{H}_b(X) = H_b(X) + \text{Tr}_s\hat{h}/N$ and $\overline{h}(R) = \hat{h}(R) - \text{Tr}_s\hat{h}(R)/N$ where $\text{Tr}_s$ denotes the trace over subsystem DOF.  The local, infinitesimal time-step analysis can be extended to a finite-interval time scale if the orthogonality approximation is not used and, in this case, results using either form of the Hamiltonian will be identical.  However, if the orthogonality approximation is made, global errors which depend the form of the Hamiltonian will accumulate. The traceless version of the Hamiltonian often yields similar or better results than the trace form, since it is less susceptible to problems due to inverted potentials. Analysis of the results of the dual avoided crossing (Tully 2) and the Fenna-Mathews-Olsen (FMO) models support this conclusion.  For consistency, we will display the Hamiltonian in trace-form when we introduce the models in this section, although calculations have been carried using both forms as indicated below.

Second, we analyze the effects of focused initial conditions on the FBTS. We use results of the Tully 2 model and spin-boson models to illustrate the trajectory picture described in Sec.~\ref{sec:focusing}.  Through these examples, we show that the Tully 2 and symmetric spin-boson models represent a small set of special cases where the use of focused initial conditions might yield results similar to those of the full sampling calculation.

Last, we consider the application of iterative focusing with JFBTS to alleviate the onerous demands on the very large number of trajectories required in the full-sampling calculations.  From the results on the asymmetric spin-boson and conical intersection models, we demonstrate the convergence of the JFBTS to the QCLE solution.  Due to the fact that trajectories collapse onto a pair of diabatic surfaces at each jump, it turns out that the selection of the subsystem basis (\textit{i.e.}, which diabatic surfaces are used) in the formalism affects the convergence of the results. We provide general guidance on how to select an ideal subsystem basis for a JFBTS calculation.

\subsection{Dual Avoided Crossing Model}

The subsystem Hamiltonian for the Tully 2 model in the diabatic basis $\{\ket{1},\ket{2}\}$ is
\begin{eqnarray}\label{eq:tully2}
\hat{h}(R) = \left[ \begin{array}{cc}
0 & Ce^{-DR^2} \\
Ce^{-DR^2} & -Ae^{-BR^2}+E_0 \
\end{array}
\right].
\end{eqnarray}
The numerical values of parameters $A$, $B$, $C$, $D$, $E_0$ and all other details of this particular model calculation are taken from Ref.~[\onlinecite{nassimi10}].  The partially Wigner transformed Hamiltonian of this system is $H_W(X) = P^2/2M + \hat{h}(R)$. Note that  the pure bath term contains only the kinetic energy of a single bath particle.  Initially, the quantum subsystem is in the state $\ket{1}$ and the bath particle is modeled as a Gaussian wave packet centered at $R_0$ with initial bath momentum $P_0$ directed towards the interaction region.

\begin{figure}[h]
     \begin{center}
     \includegraphics[width=0.9\textwidth]{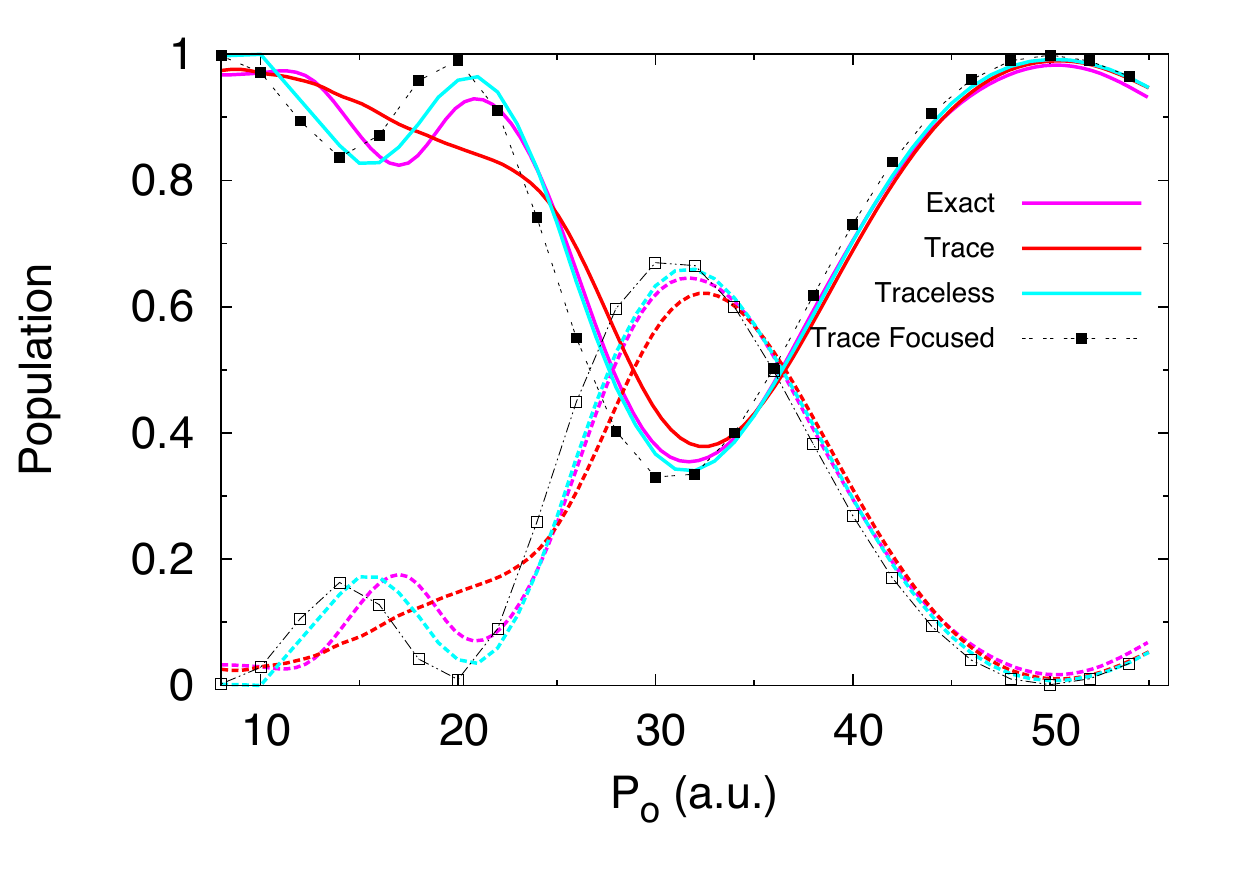}
     \end{center}
     \vspace{-20pt}
     \caption{Asymptotic populations of the diabatic state 1 (solid lines and solid squares) and state 2 (dashed lines and open squares) as functions of the initial momentum $P_0$ of the incident wave packet.}
   \label{fig:tully2}
\end{figure}

Figure~\ref{fig:tully2} shows the asymptotic populations of the quantum subsystem in the two diabatic states as a function of the initial momentum $P_0$. In this figure, we compare three different FBTS results (traceless form, trace form, trace form with focused initial conditions) with the exact quantum results. At high momentum, $P_0 > 35$, all different FBTS results converge to the exact quantum calculation.  This convergence is due to the fact that Gaussian wave packet passes through the interaction region of the scattering potential with high velocity, and the system-bath coupling does not strongly influence the quantum and bath dynamics. In the low momentum regime, the system-bath coupling plays a much more crucial role in the bath dynamics, which in turn influences the quantum dynamics.  It is clear that the traceless form (light blue curves) result match to the exact result the most.

We expect that simulations using the traceless-form of the Hamiltonian will yield the best results, due to the possible presence of an inverted potential when the trace-form Hamiltonian is used.  The fact that the Tully 2 model has no pure bath potential exacerbates the errors arising from dynamics on an inverted potential surface. It is interesting that the trace-form result improves significantly when the focusing approximation is used.  We attribute the success of the focusing approximation in this case to the validity of the adiabatic approximation in the low momentum regime (up to roughly $P_0=15$).  The probability of finding the quantum subsystem in the adiabatic ground state at any time is over $90 \%$ for initial momenta up to $P_0 = 15$.  Since focused initial conditions describe adiabatic dynamics and some additional improvements, the much more accurate low-momentum result (in comparison to the full sampling calculation with the trace-form Hamiltonian) can be understood.

Next we discuss how the FBST's formal invariance to the trace of the Hamiltonian is recovered as we introduce discontinuous jumps in the phase space.  We studied the convergence of the asymptotic population as a function of the number of jumps introduced in the JFBST calculations with both forms of the Hamiltonian.  As expected, we found that both sets (traceless and trace forms) of results improved and converged to each other.  For instance, for the case of $P_0=20$, the trace-form results converge to the exact result shown in Fig.~\ref{fig:tully2} with 4 jumps.  Similarly, the traceless-form results also converge to the exact result with 2 jumps.

From these results, we see that the traceless-form FBTS is the preferred algorithm for numerical computations.  We also see the usefulness of focused initial conditions in situations where the dynamics is weakly nonadiabatic.

\subsection{Spin-Boson Models}

The symmetric (unbiased) and asymmetric (biased) spin-boson models provide additional insight into the utility of the algorithms.  Since spin-boson models and their generalizations have been studied extensively in different sub-fields of physics and chemistry, we focus on an explanation of the performances of FBTS and JFBTS algorithms in the reproduction of exact quantum results.

The partially Wigner transformed Hamiltonian of the spin-boson model reads,
\begin{eqnarray}\label{eq:sbmodel}
\hat{H}_W(X) & = & \sum_i \left( \frac{P_i^2}{2M_i} + \frac{1}{2}M_i\omega_i^2R_i^2 - c_i R_i\hat{\sigma}_z \right) \nonumber \\
& & + \epsilon \hat{\sigma}_z - \Omega \hat{\sigma}_x,
\end{eqnarray}
where $M_i$ and $\omega_i$ are the mass and frequency of the $i$-th bath oscillator, respectively, $c_i$ controls the bilinear coupling strength between the $i$-th oscillator and the two-level quantum subsystem, $\Omega$ is the coupling strength between the two quantum levels, $\epsilon$ is the bias, and $\hat{\sigma}_{z(x)}$ is the $z (x)$ Pauli matrix.  We assume that the bilinear coupling in the spin-boson models is characterized the ohmic spectral density, $J(\omega) = \pi\sum_i c_i^2/(2M_i\omega_i)\delta(\omega-\omega_i)$, where
$c_i = (\xi\Delta M_j)^{1/2}\omega_i$, $\omega_i = -\omega_c \ln(1-i\Delta\omega/\omega_c)$, and $\Delta\omega = \omega_c(1-e^{-\omega_{max}/\omega_c})/N_B$ with $\omega_c$ the cut-off frequency, $N_B$ the number of bath oscillators, and $\xi$ the Kondo parameter.  In all the spin-boson models considered here, the quantum subsystem is initialized in state $\ket{1}$ and the bath is initially in thermal equilibrium.

\begin{figure}[h!]
     \begin{center}
     \includegraphics[width=0.9\textwidth]{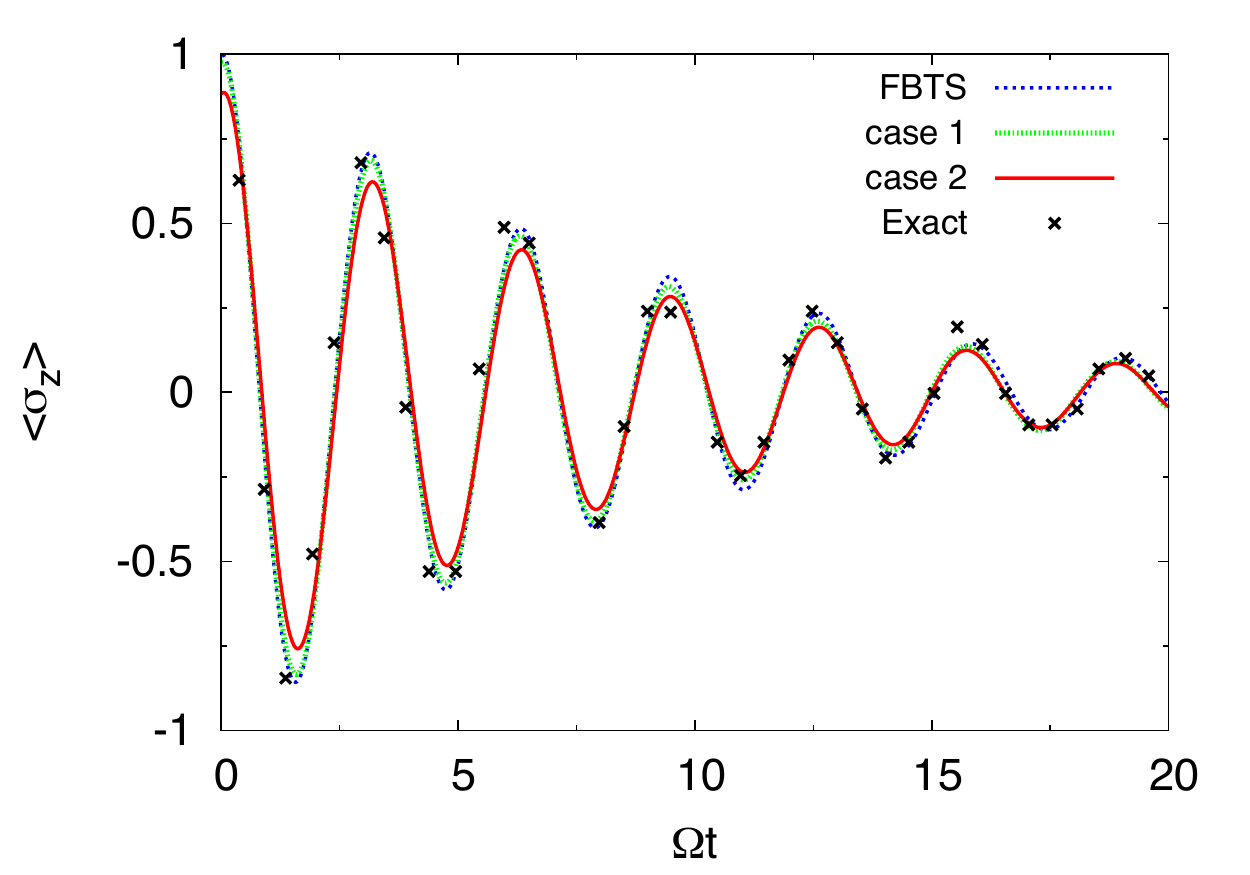}
     \end{center}
     \vspace{-17pt}
     \caption{Population difference as a function of $\Omega t$ with the parameter set: $\epsilon=0$, $\Omega=0.4$, $\xi=0.09$, $\beta=12.5$, and $\omega_c=1.0$. See the text for the two modified focused initial conditions.}
   \label{fig:symmsb}
\end{figure}

First we consider FBTS results for the symmetric spin-boson model.  Figure~\ref{fig:symmsb} shows the time evolution of the population difference, $\langle \hat{\sigma}_z \rangle$.  We not only consider the full-sampling and focused initial condition FBTS results, but also consider two cases with modified focused initial conditions.  In case 1, we replace the typical focusing condition $\delta(q_1^2+p_1^2-1)$ by $\delta(q_1^2+p_1^2-1.5)$.  In case 2, we use the even more severely modified focused initial condition $\delta(q_1^2+p_1^2-0.8)\delta(q_2^2+p_2^2-0.2)$.  In both modified cases, we omit the backward coherent states because they can be exactly replaced by the forward states as discussed in Sec.~\ref{sec:fic}.  All FBTS results in Fig.~\ref{fig:symmsb} are scaled in the following way: $<\sigma_z>/1_{MC}$, where $1_{MC}$ is the MC estimate of the constant $1$ using either the full sampling, proper or modified focused initial conditions, depending on the case considered.

Examination of Fig.~\ref{fig:symmsb} reveals a rather important reason behind the reported accounts~\cite{bonella05,kim-map08} of high accuracy in the application of focused initial conditions to the symmetric spin-boson model.  The results in the figure show that any modified focused initial condition of the form $\delta(q_1^2+p_1^2-r)$ with $r$ being an arbitrary real number can yield results (scaled by $1_{MC}$) almost identical to the exact quantum calculation.  In addition, case 2, where the modified initial condition has non-zero quantum amplitude in state $\ket{2}$, also yields good results, except that the initial amplitude is slightly lower.  This behavior, in which a large portion of phase space leads to similar quantum evolution, can be traced to the fact that the two diabatic states are energetically degenerate in the absence of the bath.

For FBTS and related mapping algorithms, it is crucial to simulate the correct bath dynamics.  We argue intuitively that the energetic degeneracy of the two states gives an effectively trivial quantum feedback on the bath. In the ensemble picture, the bath oscillators oscillate around their equilibrium points as if the quantum subsystem only influences the amplitudes of their oscillations but does not modify their oscillation frequencies or displace them from their equilibrium points. However, the quantum subsystem still feels the bath dynamics and suffers decoherence. We support this intuitive argument by the following observations:  First, if one simply suppresses the subsystem-related part in the equation of motion for P in Eq.~(\ref{eq:vol-nonH}), \textit{i.e.} use $dP/dt = -\partial V_0(R)/ \partial R$, then one still obtains a result which is almost identical to that shown in Fig.~\ref{fig:symmsb}.  Second, after the fifth oscillation in the figure, even case 2 converges to the exact result.  This suggests that once the system moves beyond its correlation time scale with respect to the initial condition, the system is insensitive to the population distribution over the two energetically degenerate levels.


The asymmetric spin-boson model in which a bias $\epsilon$ is introduced is a more challenging system for simulation.  The consequences of having a bias in the spin-boson model have been analyzed extensively; for instance, the fundamental differences between the quantum dynamics in both symmetric and asymmetric spin-boson models have been summarized in Ref.~[\onlinecite{loss05}].

\begin{figure}[ht!]
     \begin{center}
    \includegraphics[width=0.75\textwidth]{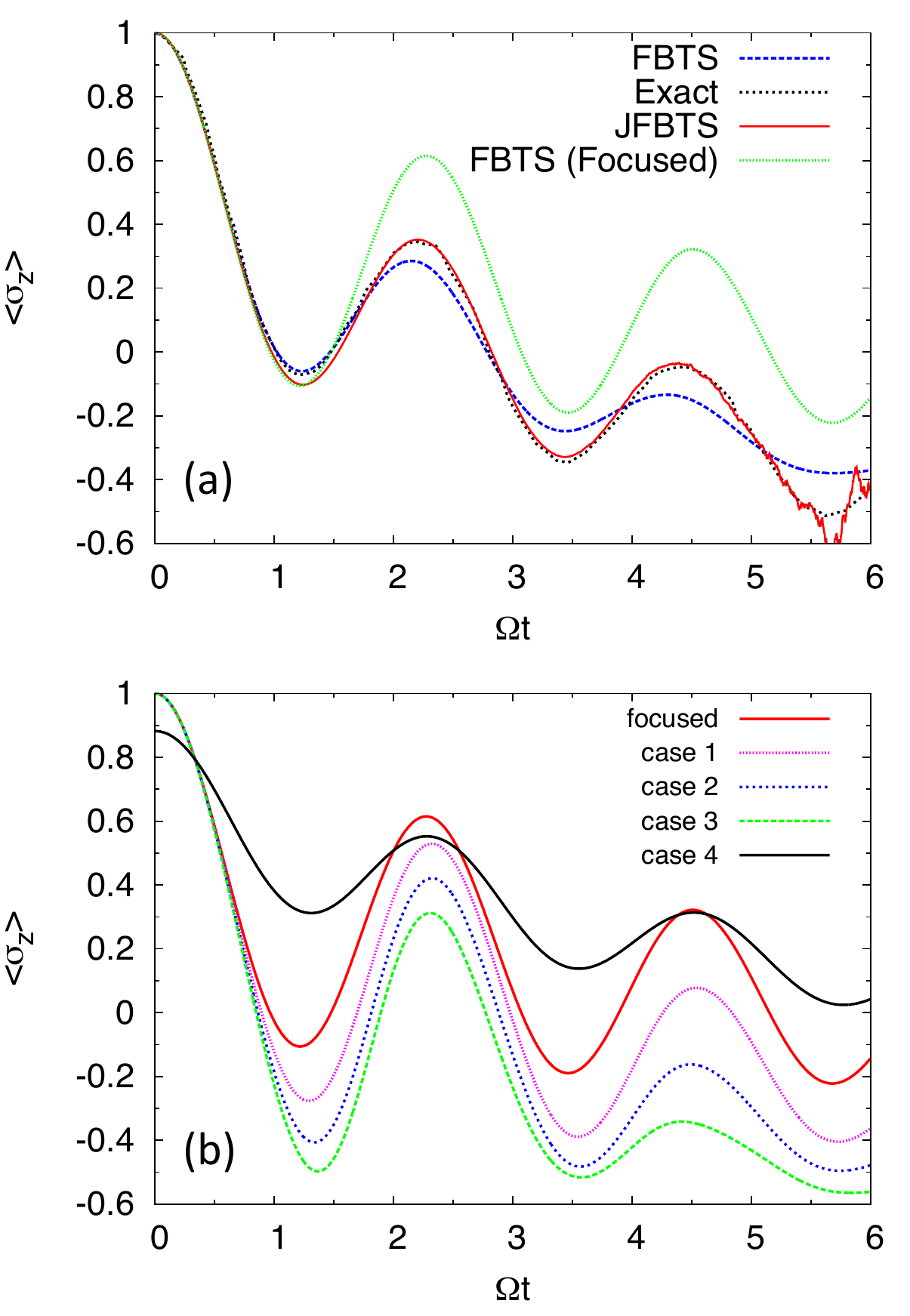}
     \end{center}
     \vspace{-20pt}
     \caption{Population difference as a function of $\Omega t$ with the following parameters: $\epsilon = \Omega = 0.4$, $\xi = 0.13$, $\beta = 12.5$ and $\omega_c = 1.0$. (a) Comparison of results obtained with the FBTS and its variants with the exact quantum values. (b) Comparison of results obtained with the FBTS with proper and modified focused initial conditions.  See text for definitions of the four cases.}
   \label{fig:asymmsb}
\end{figure}

Figure~\ref{fig:asymmsb}a presents the time evolution of $<\sigma_z>$ for several situations: FBTS (full sampling), FBTS (focused initial condition) and JFBTS (with 26 jumps).  For the asymmetric spin-boson model, we find that the focused initial condition result deviates significantly from both the full-sampling and exact quantum results, after about $\Omega t > 1.5$.  The rather poor performance of the focusing approximation to the FBTS can again be understood by analyzing the trajectories with modified focused initial conditions.  In Fig.~\ref{fig:asymmsb}b, several such cases are presented.  For cases 1-3, the modified initial conditions take the form $\delta(q_1^2+p_1^2-r)$ with $r=1.5$, $1.8$, and $2.0$, respectively.  In case 4, we use the initial condition $\delta(q_1^2+p_1^2-0.8)\delta(q_2^2+p_2^2-0.2)$. All the results in Fig.~\ref{fig:asymmsb}b are scaled by the corresponding $1_{MC}$ as explained earlier.  When  $\epsilon \neq 0$, different initial conditions yield very different results for $\langle \sigma_z \rangle$.  Comparing Figs.~\ref{fig:asymmsb}a and b, it is clear that the proper focused initial condition is simply not a valid approximation since it prunes too many trajectories from the ensemble.

In Fig.~\ref{fig:asymmsb}a, we also present results for a JFBTS simulation with 26 jumps.
We remark that the minimum number of average jumps required for a JFBTS calculation to recover the QCLE result depends on the interplay of several factors, such as the size of the time-step, the probability distribution $P_{\{\kappa\}}$ and the time-block size J discussed in Sec.~\ref{sec:jfb}.  In this calculation, we used the binomial distribution for $P_{\{\kappa\}}$, and a jump could be introduced in every even-number time step.  Finally, we remark that, in certain cases,  the QCLE result might be obtained with fewer jumps if one uses fixed-interval jumps. In summary, the JFBTS formalism is flexible in the sense that many parameters and the jump process can be modified in order to optimize its performance for each individual problem.

From the analysis of these spin-boson models, we see that the applicability of focused initial conditions is very limited in scope.  Their ability to describe symmetric spin-boson cases should be taken simply as a rare exception.  However, the use of iterative focusing in the JFBTS was shown to be a more robust and useful tool.  It allows us to obtain the QCLE solution, which agrees with the exact quantum result.

\subsection{Fenna-Matthews-Olson Model}

The Fenna-Matthews-Olson (FMO) protein of green sulfur bacteria plays a role in the transfer of excitation energy to the reaction center involved in photosynthesis~\cite{fleming09}. The model Hamiltonian for this process provides an example of a multi-level quantum system coupled to a bath.  The Hamiltonian is composed of a seven-level quantum subsystem corresponding to an excitation state localized in one of the seven pigment proteins of the FMO complex, and each quantum level is bilinearly coupled to its own set of bath oscillators; the bilinear system-bath coupling is characterized by the Debye spectral density. The explicit expression for the Hamiltonian and the parameters used in its definition will not be given here but is available in the Ref.~[\onlinecite{fleming09a}] and its supporting documents online.  The quantum subsystem is initialized in the state $\ket{1}$ and all the bath oscillators are initially in the thermal equilibrium state. This model has been studied often.  In particular, the PBME~\cite{kelly-FMO} and the PLDM~\cite{huo11} algorithms, closely related to the FBTS, have already been applied to investigate the population dynamics of the FMO complex and compared to the numerically accurate quantum results~\cite{fleming09a, zhu11}.

\begin{figure}[h!]
     \begin{center}
     \includegraphics[width=0.9\textwidth]{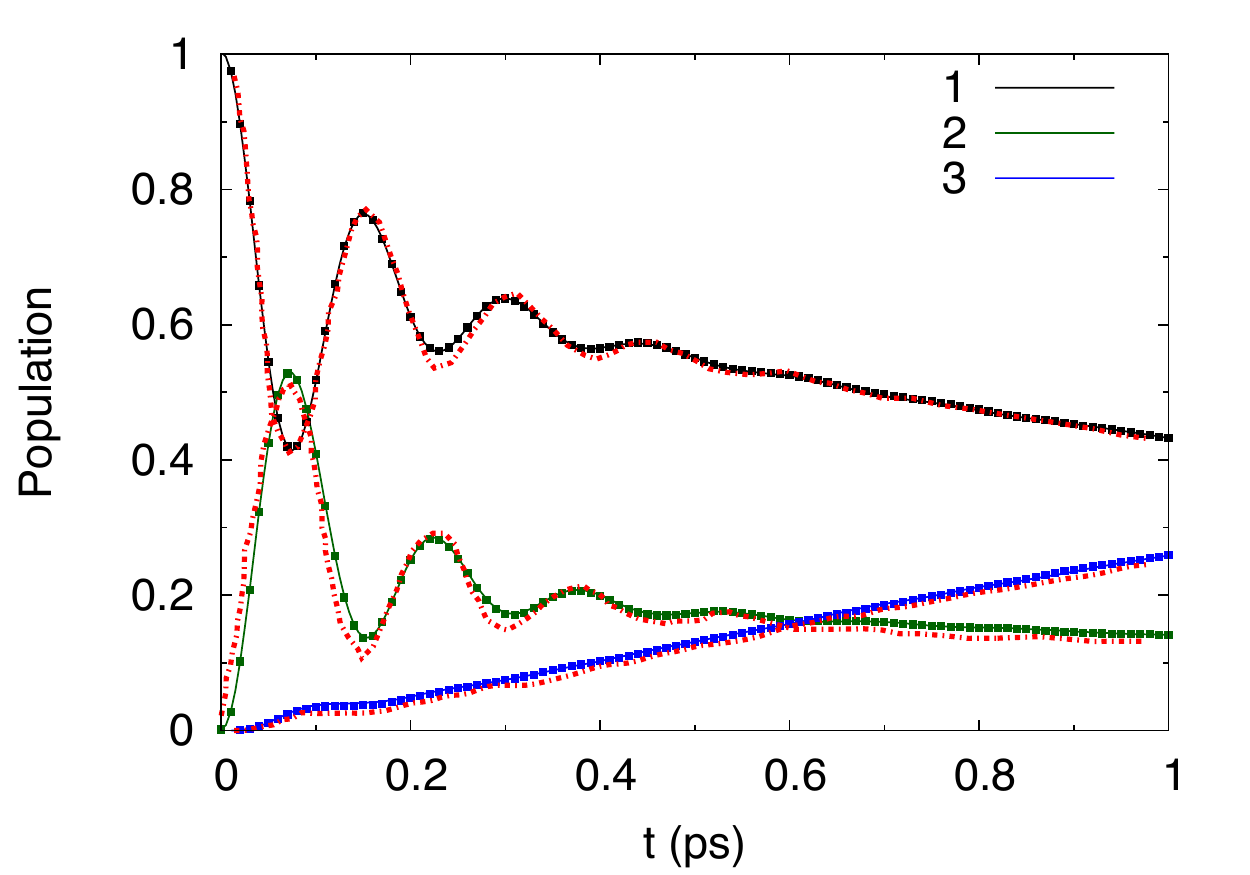}
     \end{center}
     \vspace{-20pt}
     \caption{Population distribution of bacteriochlorophyll (Bchl.) 1-3 as function of $t$ at temperature of 77 K. The solid lines represent the traceless-form results, and the corresponding color dots represent the trace-form results.  The red data points are extracted from Ref.~[\onlinecite{zhu11}] }.
   \label{fig:fmo}
\end{figure}

Figure~\ref{fig:fmo} plots the populations in states $\ket{1}$, $\ket{2}$, and $ \ket{3}$ as functions of time.  Two sets of results are presented and compared against quantum results obtained from rescaled Hierarchical Coupled Reduced Master Equation (HCRME) calculation~\cite{zhu11}.  The solid lines represent the results of the traceless-form FBTS, while the corresponding colored dots correspond to results of the trace-form FBTS.  For this system the FBTS is rather insensitive to whether the trace or traceless forms of the Hamiltonian are used.  Unlike the Tully 2 model, the FMO model has a dominant pure bath quadratic potential for each oscillator.  Since the system-bath coupling is a weak perturbation, it is less likely to encounter an inverted potential for the bath trajectories. In addition to the agreement between the trace-form and traceless-form, the results are also in quantitative accord with the rescaled HCRME calculations~\cite{zhu11}. Furthermore, we have also checked and found (not shown) that the long-time population distributions of these states approach the thermal equilibrium distribution at time $t=10$ ps.

The results of this section demonstrate the efficiency of the FBTS in dealing with multi-level systems.  For this system, inverted potentials in the simulation of trajectories do not present a problem and the FBTS is not sensitive to the form of the Hamiltonian.

\subsection{Conical intersection model}

Finally we consider a two-level, two-mode quantum model for the coupled vibronic states of a linear ABA triatomic molecule constructed by Ferretti, Lami and Villiani (FLV).  In this model, the nuclei are described by two vibrational degrees of freedom, $X$ and $Y$.  The partially Wigner transformed Hamiltonian reads
\begin{eqnarray}\label{eq:flv_bath}
H_W(R_s,P_s) & = & \frac{P_X^2}{2M_X}+\frac{P_Y^2}{2M_Y}+\frac{1}{2}M_Y\omega_Y^2Y^2 +\hat{h}(R_s), \nonumber \\
\end{eqnarray}
where the subsystem Hamiltonian is defined by the following matrix elements:
\begin{eqnarray}\label{eq:flv_qm}
h^{11}(R_s) & = & \frac{1}{2} M_X \omega_X^2 (X-X_1)^2, \nonumber \\
h^{22}(R_s) & = & \frac{1}{2} M_X \omega_X^2 (X-X_2)^2 + \Delta, \nonumber \\
h^{12}(R_s) & = & \gamma Y \exp\left(-\alpha(X-X_3)^2-\beta Y^2\right).
\end{eqnarray}
In these equations, $R_s=(X,Y)$, $P_s=(P_X,P_Y)$, $M_{X,Y}$ and $\omega_{X,Y}$ are the mass and frequency for the $X$ and $Y$ DOF, respectively. The quantum subsystem is initialized in the adiabatic ground state, while the vibronic $X$ and $Y$ initial states are taken to be Gaussian wave packets.
Further details of this model can be found in Refs.~[\onlinecite{kelly10}] and [\onlinecite{ferretti_1}].

\begin{figure}[h!]
     \begin{center}
     \includegraphics[width=0.9\textwidth]{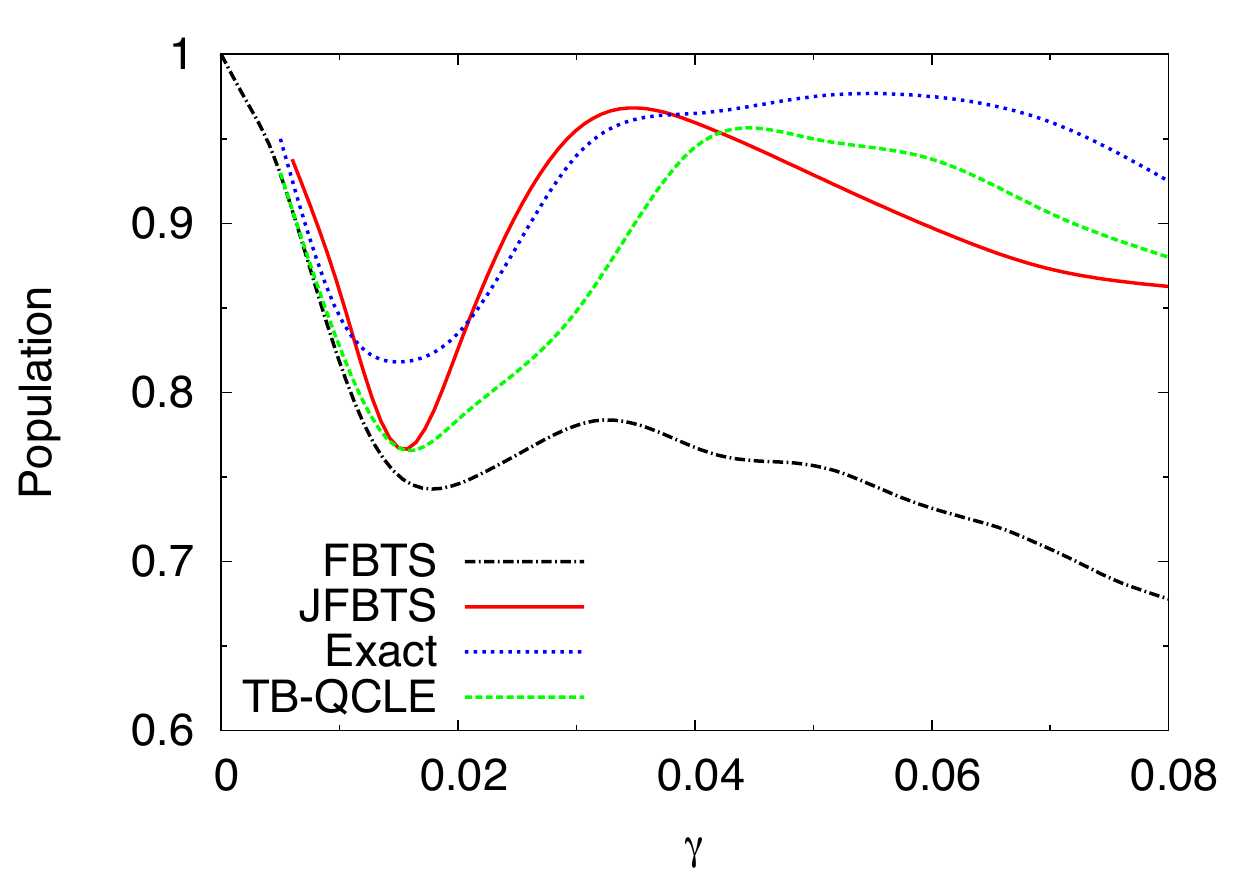}
     \end{center}
     \vspace{-17pt}
     \caption{Asymptotic adiabatic ground state population at 50~fs versus $\gamma$.  Due to the strong system-bath coupling, the orthogonality approximation compromises the FBTS results.}
   \label{fig:flv_gamma}
\end{figure}

Figure~\ref{fig:flv_gamma} plots the adiabatic ground state population at $t=50$ fs as a function of the coupling strength $\gamma$.  We see that the FBTS result matches the Trotter-based QCLE (with filtering) result up to about $\gamma = 0.02$ but deviates significantly from the QCLE and exact quantum results for larger coupling strengths.  At higher coupling strengths, the errors induced by the orthogonality approximation become significant.  The result of a 15-jump JFBTS calculation is also shown.  This simulation is able to reproduce all trends in population versus coupling strength curve and, for this number of jumps, about $10~\%$ accuracy is achieved at the highest coupling.


Next we discuss the importance of the subsystem basis used for the JFBTS calculation.  The JFBTS result in Fig.~\ref{fig:flv_gamma} was obtained using a representation with the basis states $\ket{\pm} = \left(\ket{1} \pm \ket{2}\right)/\sqrt{2}$ where subsystem states $\ket{1}$ and $\ket{2}$ were used to define the Hamiltonian matrix elements in Eq.~(\ref{eq:flv_qm}).  In the figure, this rotated basis yields a JFBTS result with as much as $\sim 25 \%$ improvement over the FBTS results. In our other JFBTS calculation (not shown in the figure) with respect to the original basis, $\{ \ket{1} , \ket{2} \}$, we obtained at most $\sim 15 \%$ improvement over the FBTS results with the same number of jumps.

As discussed in Sec.~\ref{sec:if}, the importance sampling used to select a pair of forward and backward state indices, $(s_i, s'_i)$, for the collapse of the trajectories in each MC run helps to avoid doing the exact summation over state indices in Eq.~(\ref{eq:mqc_soln-jump2}).
Thus, another key to the efficiency of the JFBTS algorithm is to have a dispersed probability distribution for $P_{s_i,s'_i}$ in order to select as many different combinations of states as possible.  An easy method to satisfy this requirement is to simply use the uniform probability distribution $P_{s_i,s'_i} = 1/N^2$ where $N$ is the number of quantum states.  However, we prefer the distribution $P_{s_{i},s^{\prime}_{i}} = (q_{s_i}^2+p_{s_i}^2)(q^{\prime 2}_{s^{\prime}_{i}}+p^{\prime 2}_{s^{\prime}_{i}})/\mathcal{N}_i$ defined in Sec.~\ref{sec:if} since it minimizes the abrupt changes that can influence the bath dynamics. Another way to obtain a more uniform distribution is to select a basis where the off-diagonal matrix elements are maximized since these influence population transfer among different levels.

\begin{figure}[h!]
     \begin{center}
    \includegraphics[width=0.9\textwidth]{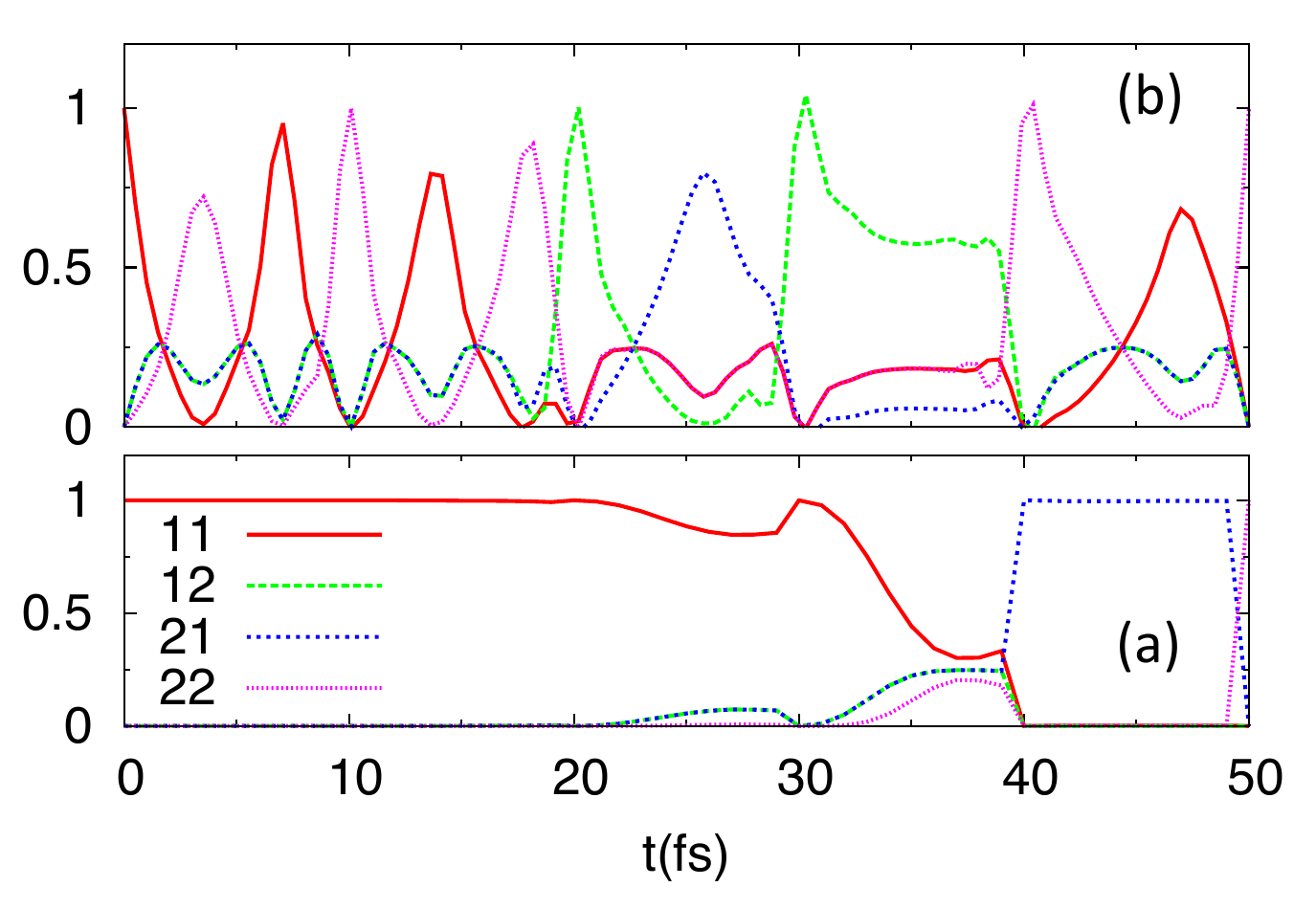}
     \end{center}
     \vspace{-15pt}
     \caption{The probability distribution $P_{s,s'}$ for importance sampling of a pair of forward and backward quantum states versus $t$ for a typical trajectory in the original basis (a) and rotated basis (b). In the later case, a more dispersed probability distribution allows for a more balanced samplings of all pairs of state combinations. One should replace 1/2 with +/- when using the legend in panel a to interpret the curves in panel b .}
   \label{fig:flv_mtx}
\end{figure}

In the original basis, the FLV model has a sharply peaked off-diagonal matrix element, $h^{12}(R_s)$.  Thus, in the asymptotic region where $h^{12}(R_s) \approx 0$, the population distribution over the quantum states does not change significantly even if jumps are introduced in the JFBTS calculation. Figure~\ref{fig:flv_mtx}a shows the population distribution over pairs of forward and backward quantum states for a typical trajectory in the original basis.  The distribution is highly localized on one particular pair of states, except when the system moves through the interaction region where $h^{12}(R_s)$ approaches maximum strength; however, $h^{12}(R_s)$ has a rather steep profile and the interaction region occupies a very small portion of the phase space.  Therefore, extensive MC sampling involving many jumps would be required to sample all different pairs of states.  By contrast, in the rotated basis $h^{++, (--)}(R_s) = \pm h^{12}$ are now the diagonal elements of the Hamiltonian matrix, and the off-diagonal matrix element $h^{+-}(R_s)$ is linearly dependent on $X$.   In the ``asymptotic'' region (where $h^{12}(R_s) \approx 0$), the two diagonal matrix elements are approximately energetically degenerate in the rotated basis. Consequently, population transfer occurs easily in the ``asymptotic'' region.  Since the ``asymptotic'' region represents the bulk of the phase space, Fig.~\ref{fig:flv_mtx}b shows a more balanced and oscillating population distribution over pairs of forward and backward quantum states for a typical trajectory in the rotated basis.

These results show that the FBTS formalism can be customized and several modifications, such as the choice of subsystem basis and the jump probability distribution, etc., can be independently tuned and adapted to suit the problem under study without actually changing the structure of the algorithm.  From the analysis of FLV model, we have seen how to determine the most suitable subsystem basis for the JFBTS calculation.


\section{Summary and Conclusion} \label{sec:con}
In this work, we demonstrated the usefulness of the FBTS through simulations of a variety of systems.  In most instances, we obtained converged results with $\sim 10^4$ trajectories.  These numerical tests suggest that the FBTS performs at the same level of accuracy and efficiency as the PLDM\cite{huo11,huo12_mp,huo12_jcp}, a closely related algorithm.   We showed that the traceless-form Hamiltonian yields more accurate results when there is no dominant pure bath potential in the model and should be used for all numerical computations.  In cases where the FBTS results were compromised by the orthogonally approximation,  we demonstrated that these results could be systematically improved by increasing the number of jumps in the generalized JFBTS algorithm.  In order to make JFBTS a computationally efficient algorithm, we carried out a detailed investigation of the focusing approximation and provided an extensive analysis of the approximate trajectory dynamics.  This analysis indicated that focused initial conditions should be used only when the nonadiabatic effects can be treated as a weak perturbation or for special cases such as the symmetric spin-boson model where the bath is rather insensitive to the differences between the quantum states.  Finally, we also justified the validity of iterative focusing in the JFBTS. In all cases studied the QCLE solution either reproduced the exact quantum results or was a very good approximation to the exact results. Thus, the ability of the FBTS and JFBTS to reproduce the QCLE solution also implies an ability to reproduce the full quantum results.

The number of jumps used in the JFBTS determines how many trajectories will be needed to obtain a converged result.  As noted throughout the paper, several implementation details of the JFBTS algorithm can be customized to significantly reduce the minimum number of jumps and improve the convergence properties of the algorithm.  In addition, the number of jumps needed to obtain QCLE result will certainly depend on the details of the Hamiltonian of each model.  It is not a simple manner to predict the number of jumps and trajectories required to obtain the full QCLE result for an arbitrary system.   However, at least two orders of magnitude increase in the number of trajectories (in comparison to the corresponding FBTS calculation) should be expected for most cases in which the FBTS results are compromised by the orthogonality approximation.  Perhaps the greatest utility of the JFBTS algorithm is that it provides a simple method to gauge the sufficiency of the FBTS when one studies nonadiabatic dynamics for complex real systems, beyond the simple benchmark models studied here for which exact quantum results are known.

Potential improvements to the family of (J)FBTS algorithms are still possible; especially, it might be helpful to reformulate the entire formalism in the adiabatic basis.  Currently, the flexibility of the algorithm can be traced back to the use of the subsystem basis, which is not uniquely defined as in the case of adiabatic basis.  In particular, a current challenge of the JFBTS algorithm is related to the determination of the best timings for the insertion of jumps in the course of trajectory evolution. In the current formulation, there are no indicators which help to make this judgement.  However, as is often the case with surface hopping algorithms formulated in the adiabatic basis, we expect that inherent indicators naturally emerge from the formalism once it is re-expressed in the adiabatic basis.

\begin{acknowledgments}
This work was supported in part by a grant from the Natural Sciences and Engineering Council of Canada.  Part of the computations were performed on the general purpose cluster supercomputer at the SciNet HPC Consortium. SciNet is funded by the Canada Foundation for Innovation under the auspices of Compute Canada, the Government of Ontario, Ontario Research Fund-Research Excellence, and the University of Toronto.
\end{acknowledgments}

\appendix
\section{Insertions of Projection Operators }

In this appendix we support the claim that insertions of projection operators, $\mathcal{P} = \sum \ket{m_s}\bra{m_s}$, in the mapping representation do not introduce additional errors.  We first consider the concatenation of parameterized time evolution operators in the subsystem:
\begin{eqnarray}
				&& \langle \lambda \vert e^{-\frac{i}{\hbar}\hat{h}(X_0)\tau} \dots e^{-\frac{i}{\hbar}\hat{h}(X_{M-1})\tau} \vert \lambda' \rangle \nonumber \\
				&& = \quad \sum_{\mu_1,\dots, \mu_{M-1}} \langle \lambda \vert  e^{-\frac{i}{\hbar}\hat{h}(X_0)\tau} \vert \mu_1 \rangle \langle \mu_1 \vert
	 e^{-\frac{i}{\hbar}\hat{h}(X_1)\tau} \vert \mu_2 \rangle \nonumber \\
	 && \qquad \quad \times \dots \langle \mu_{M} \vert  e^{-\frac{i}{\hbar}\hat{h}(X_{M-1})\tau} \vert \lambda' \rangle \nonumber \\
	 && = \quad \sum_{\mu_1,\dots, \mu_{M-1}} \langle m_\lambda \vert  e^{-\frac{i}{\hbar}\hat{h}_m(X_0)\tau} \vert m_{\mu_1} \rangle \nonumber \\
     && \qquad \quad \times \langle m_{\mu_1} \vert
	 e^{-\frac{i}{\hbar}\hat{h}_m(X_1)\tau} \vert m_{\mu_2} \rangle \nonumber \\
	 && \qquad \quad \times \dots \langle m_{\mu_{M}} \vert  e^{-\frac{i}{\hbar}\hat{h}_m(X_{M-1})\tau} \vert m_{\lambda'} \rangle.
\end{eqnarray}
On the right side of the first equality, we insert the resolution of identity $\mathcal{I} = \sum_\mu \ket{\mu}\bra{\mu}$ at every time step.  To obtain the second equality, we use the matrix identity: $\bra{\mu} e^{\frac{i}{\hbar}\hat{h}(X)\tau} \ket{\mu'} = \bra{m_\mu} e^{\frac{i}{\hbar}\hat{h}_m(X)\tau} \ket{m_{\mu'}}$.  If we express every time evolution operator in terms of $z_i$ variables,
\begin{equation}
e^{\frac{i}{\hbar}\hat{h}_m(X)\tau} = \int \frac{d^2 z}{\pi^N} \ket{z} \bra{z(\tau)},
\end{equation}
where $\bra{z(\tau)}$ is time-evolved by Eq.~(\ref{eq:vol-nonH}), we obtain the desired expressions in Eq.~(\ref{eq:mqc_soln-jump}).

\section{Forward and Backward Trajectories with Focused Initial Conditions}

In this appendix we take $\hbar = 1$ and use the scaled coherent state variables as defined in Sec.~\ref{sec:theory}.
We prove that the backward trajectories can be exactly replaced by the forward trajectories in the FBTS if the focused initial condition is imposed.  The coherent state variables, $z$ and $z'$, matter critically in two places in the formalism.  First, the equation of motion for the bath momenta defined in Eq.~(\ref{eq:vol-nonH}) depends explicitly on these variables through the term $\frac{1}{2}\frac{\partial h^{\lambda\lambda'}(R)}{\partial R}(q_\lambda q_{\lambda'} + p_\lambda p_{\lambda'} + q'_{\lambda} q'_{\lambda'} + p'_{\lambda}p'_{\lambda'})$.  Second, these variables play a critical role in the evaluation of the expectation value of an operator $\hat{B}$,
\begin{eqnarray}
				\langle \hat{B}(t) \rangle & = & \sum_{\lambda,\lambda'} \int dX \int dx dx' B^{\lambda'\lambda}_W(X,t)\rho_W^{\lambda\lambda'}(X), \nonumber \\
				& = & \sum_{\substack{\lambda,\lambda'\\\mu,\mu'}} \int dX \int dx dx' \phi(x) \phi(x') \rho^{\lambda\lambda'}_W(X) \nonumber \\
				& & \times (q_\lambda - ip_\lambda)(q_\mu(t) + ip_\mu(t))\\
				& & \times (q'_{\lambda'}+ip'_{\lambda'})(q'_{\mu'}(t)-ip'_{\mu'}(t))B_W^{\mu\mu'}(X_t),\nonumber
\end{eqnarray}
where the $q$ and $p$ variables with no explicit time arguments are taken to be the initial values at $t=0$.
In the following, we prove the identities:
\begin{eqnarray}\label{eq:fwbwidenty}
& & q'_{\lambda}(t) q'_{\lambda'}(t) + p'_{\lambda}(t) p'_{\lambda'}(t)=q_\lambda(t) q_{\lambda'}(t) + p_\lambda(t) p_{\lambda'}(t),\\
& & (q'_{\lambda'}+ip'_{\lambda'})(q'_{\mu'}(t)-ip'_{\mu'}(t))=(q_{\lambda'}+ip_{\lambda'})(q_{\mu'}(t)-ip_{\mu'}(t)).  \nonumber
\end{eqnarray}
The first identity implies that the backward trajectories can be replaced by the forward ones in the equation of motion for the bath.  The second identity implies that the expectation value of the operator $\hat{B}(t)$ can also be evaluated without having the backward trajectories.

Without loss of generality, we assume the quantum state is initialized in state $|1 \rangle$ with the following focused initial condition:
$\delta(q_1^2+p_1^2-1)\delta(q^{\prime 2}_1 + p^{\prime 2}_1 - 1)$ and every other component of the $q$ and $p$ vectors is set to zero at $ t=0$.  We remind that the $q(t)$ and $p(t)$ variables satisfy the equation of motion in Eq.~(\ref{eq:vol-nonH}).  First we define a function $g_{\lambda\lambda'}(t) = q'_{\lambda}(t) q'_{\lambda'}(t) + p'_{\lambda}(t) p'_{\lambda'}(t)-q_\lambda(t) q_{\lambda'}(t) - p_\lambda(t) p_{\lambda'}(t)$.  It is easy to verify that $g_{\lambda\lambda'}(0) = 0$ for every $(\lambda, \lambda')$ combination.  Now we prove that $\left. \frac{d^{n} g_{\lambda\lambda'}(t)}{dt^n}\right\vert_{t=0} = 0 $ for all $n$. We show that the first two derivatives of $g_{\lambda\lambda'}(t)$ are equal to zero:
\begin{eqnarray}\label{eq:dg1}
				\left. \frac{d g_{\lambda\lambda'}(t)}{dt} \right \vert_{t=0} & = & h^{\lambda\alpha}[p'_\alpha q'_{\lambda'} - q'_\alpha p'_{\lambda'} - p_\alpha q_{\lambda'} + q_\alpha p_{\lambda'} ] \vert_{t=0} \nonumber \\
				& & + h^{\alpha\lambda'}[p'_\lambda q'_{\alpha} - q'_\lambda p'_{\alpha} - p_\lambda q_{\alpha} + q_\lambda p_{\alpha} ] \vert_{t=0} \nonumber \\
				& =  & 0,
\end{eqnarray}
and
\begin{eqnarray}\label{eq:dg2}
				\left. \frac{d^2 g_{\lambda\lambda'}(t)}{dt^2} \right \vert_{t=0} & = & 2h^{\lambda\alpha}h^{\lambda^{\prime}\beta}g_{\alpha\beta}(0)- h^{\lambda^{\prime}\alpha}h^{\alpha\beta}g_{\beta\lambda}(0)  - h^{\lambda\alpha}h^{\alpha\beta}g_{\beta\lambda^{\prime}}(0) \nonumber \\
				& =  & 0.
\end{eqnarray}
Since any higher derivative of $g_{\lambda\lambda'}(t)$ at $t=0$ can be recursively defined in terms of Eqs.~(\ref{eq:dg1}) and (\ref{eq:dg2}), we have proved that $g_{\lambda\lambda'}(t) = 0$ for all $t$ and established the first identity in Eq.~(\ref{eq:fwbwidenty}).

Next we define functions $f_{\lambda\lambda'}(t) = q'_\lambda q'_{\lambda'}(t) + p'_\lambda p'_{\lambda'}(t) - q_\lambda q_{\lambda'}(t) - p_\lambda p_{\lambda'}(t)$ and $k_{\lambda\lambda'}(t)= -q'_\lambda p'_{\lambda'}(t) + p'_\lambda q'_{\lambda'}(t) - p_\lambda q_{\lambda'}(t) + q_\lambda p_{\lambda'}(t)$. We remark that the second identity in Eq.~(\ref{eq:fwbwidenty}) is equivalent to $f_{\lambda\lambda'}(t) + ik_{\lambda\lambda'}(t) = 0$.   Similar to the previous analysis, we note that $f_{\lambda\lambda'}(t) = 0$ and $k_{\lambda\lambda'}(t)=0$ by showing that these functions and all orders of their derivatives are zero at $t=0$.  Furthermore, we note the simple relations between the first derivatives of these functions:
\begin{eqnarray}
				\left. \frac{d f_{\lambda\lambda'}(t)}{d t}\right \vert_{t=0} & = & h^{\lambda'\alpha} [q'_{\lambda}p'_{\alpha}(t) - p'_{\lambda}q'_{\alpha}(t) - q_{\lambda}p_{\alpha}(t) + p_\lambda q_\alpha(t)] \vert_{t=0}, \nonumber \\
				& = & -h^{\lambda'\alpha}k_{\lambda\alpha}(0) \nonumber \\
				& = & 0, \nonumber \\
				\left. \frac{d k_{\lambda\lambda'}(t)}{d t}\right \vert_{t=0} & = & h^{\lambda'\alpha} [ q'_{\lambda}q'_{\alpha}(t) +p'_{\lambda}p'_{\alpha}(t) - p_{\lambda}p_{\alpha}(t) - q_\lambda q_\alpha(t)] \vert_{0}, \nonumber \\
				& = & h^{\lambda'\alpha}f_{\lambda\alpha}(0) \nonumber \\
				& = & 0.
\end{eqnarray}
From these equations, it is obvious that all higher derivatives of these functions evaluated at $t=0$ should vanish.  Thus, we have established the second identity in Eq.~(\ref{eq:fwbwidenty}).



%

\newpage

\end{document}